\documentclass[review]{elsarticle}
\usepackage[T1]{fontenc}
\usepackage{dfadobe}  

\usepackage{cite}  

\usepackage{amssymb}
\usepackage{latexsym}

\usepackage[]{algorithm2e}
\usepackage{color}
\usepackage{xcolor}
\usepackage{xspace}
\usepackage{tabularx}

\usepackage{changes}
\definechangesauthor[name={Daniele Prada}, color=blue]{dp}

\newcommand{\Pinabla}{\Pi^\nabla}
\newcommand{\Svem}{S}

\newcommand{\Th}{\mathcal{T}_h}
\newcommand{\Eh}{\mathcal{E}_h}
\newcommand{\EK}{\mathcal{E}^K}

\newtheorem{theorem}{Theorem}[section]
\newtheorem{problem}[theorem]{Problem}
\newtheorem{assumption}[theorem]{Assumption}

\usepackage{appendix}

\title{Benchmark of Polygon Quality Metrics\\for Polytopal Element Methods}

\author{M. Attene, S. Biasotti, S. Bertoluzza, D. Cabiddu, M. Livesu,\\ G. Patan\`e, M. Pennacchio, D. Prada, M. Spagnuolo}
\ead{patane@ge.imati.cnr.it}
\address{Consiglio Nazionale delle Ricerche\\
Istituto di Matematica Applicata e Tecnologie Informatiche\\
Italy}

\begin{document}


\begin{frontmatter}
\begin{abstract}
Polytopal Element Methods (PEM) allow to solve differential equations on general polygonal and polyhedral grids, potentially offering great flexibility to mesh generation algorithms. Differently from classical finite element methods, where the relation between the geometric properties of the mesh and the performances of the solver are well known, the characterization of a good polytopal element is still subject to ongoing research. Current shape regularity criteria are quite restrictive, and greatly limit the set of valid meshes. Nevertheless, numerical experiments revealed that PEM solvers can perform well on meshes that are far outside the strict boundaries imposed by the current theory, suggesting that the real capabilities of these methods are much higher. In this work, we propose a benchmark to study the correlation between general 2D polygonal meshes and PEM solvers. The benchmark aims to explore the space of 2D polygonal meshes and polygonal quality metrics, in order to identify weaker shape-regularity criteria under which the considered methods can reliably work. The proposed tool is quite general, and can be potentially used to study any PEM solver.
Besides discussing the basics of the benchmark, in the second part of the paper we demonstrate its application 
on a representative member of the PEM family, namely the Virtual Element Method, also discussing our findings.

%

\end{abstract}  
\end{frontmatter}



\newcommand{\cino}     			[1]{{\color{magenta}			Cino: #1}}
\newcommand{\cabid}			[1]{{\color{magenta}			Cabid: #1}}
\newcommand{\michi}			[1]{{\color{magenta}			michi: #1}}
\newcommand{\marco}		    [1]{{\color{magenta}			marco: #1}}
\newcommand{\beppe}		    [1]{{\color{magenta}			beppe: #1}}
\newcommand{\bias}		    	[1]{{\color{magenta}			SBias: #1}}
\newcommand{\bert}		    	[1]{{\color{magenta}			SBert: #1}}
\newcommand{\prada}		    [1]{{\color{magenta}			Dprada: #1}}
\newcommand{\micol}		    [1]{{\color{magenta}			Micol: #1}}

\newcommand{\com} 				[1]{{}} 
\newcommand{\edit}				[1]{{\color{red}			#1}} 

\newcommand{\SG}{{\Gamma_{\mathsf{geom}}}} 
\newcommand{\SP}{{\Gamma_{\mathsf{pde}}}}  
\newcommand{\SM}{{\Gamma_{\mathsf{mesh}}}} 



\section{Introduction}
\label{sec:intro}
Solving a PDE on geometrically complex domains is a fundamental task
of scientific computing. In real applications, the generation of a good discretization of the domain is key to obtain quality results, but it can also be an extremely complex problem, often even harder than the 
numerical solution of the discretized equations~\citep{hughes2005isogeometric}. 
To alleviate meshing issues, recent literature widely explored Polytopal Element Methods (PEM), that is,
methods for the numerical solution of PDEs based on polygonal and polyhedral grids. PEM 
approaches allow to:
(i) achieve high flexibility in the treatment of complex geometries;
(ii) incorporate complex features at different scales without triggering mesh refinement; 
(iii) automatically include hanging nodes (i.e., T-junctions); (iv) simplify refinement and coarsening operators. 

Similarly to standard Finite Elements, the performance of PEM depends on the quality of the underlying mesh, in terms of accuracy, stability, and effectiveness of preconditioning techniques. However, while the concept of shape regularity of classical finite elements
is well understood~\citep{shewchuk2002good,ciarlet2002,BRANDTS20082227}, the characterization of a good polytopal element is still subject to ongoing research. 

\begin{figure}[t]
\centering
\includegraphics[width=\columnwidth]{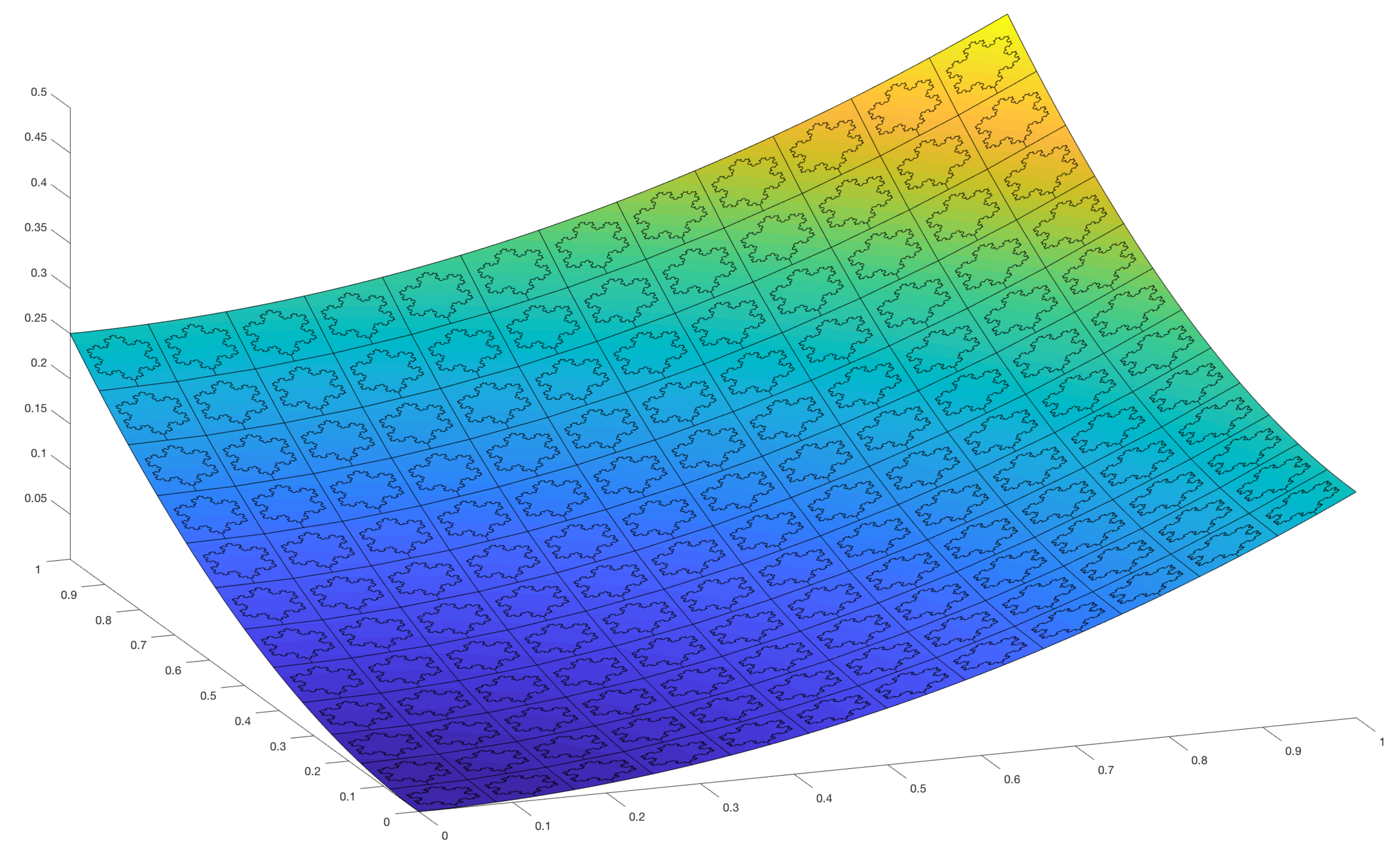}
\caption{A PDE solved on a mesh made of snowflake-like elements (and their dual, that is elements with snowflake holes). Despite the accuracy of the solution, these elements clearly violate any known shape regularity criterion for polygonal elements.}
\label{fig:snowflake}
\end{figure}

Recent works on shape regularity for PEM~\citep{shapereg,mu2015shape} require each element in the mesh to be star-shaped. Despite these tight constraints, numerical tests revealed that PEM
allows to reliably solve PDEs even on meshes made of bizarre non starred polygons, such as the ones shown in Figure~\ref{fig:snowflake}. This suggests that there is a gap between the current theory and the real capabilities of the solver, which promise to be much higher (thus allowing for much weaker requirements on the tessellation). However, unlike classical finite elements, the enormous freedom in terms of different shapes given by the polytopal framework, makes it difficult to pinpoint exactly what geometric features
do or do not have a negative effect on the performance of the method.

In this paper, we introduce 
a tool that allows to systematically explore the relation between the geometric properties of the mesh and
the performances of PEM solvers. 
Our goal is twofold: first, we aim to identify more permissive shape-regularity criteria under which the considered method is effective; second, we wish to single out specific issues that negatively affect the results, with the final aim of designing better methods. More precisely, we propose a benchmark that studies the correlation between the performance of a given PEM solver, and a wide set of polygon quality metrics. 

We considered a large set of geometric properties of polygons, spanning from areas, angles and edge lengths, to kernels, inscribed and circumscribed circles. The whole set is reported in Table~\ref{tab:geom_metrics}, and comprises 12 per polygon metrics, which become 36 per mesh metrics considering minima, maxima, and averages. 
Concerning polygonal meshes, available Voronoi based meshing tools (e.g.~\citep{du1999centroidal}) are not suited to aid our study, because they produce convex elements which are not challenging enough to stress PEM solvers. Instead, we opted for a family of parametric elements explicitly designed to progressively stress one or more of the metrics listed in Table~\ref{tab:geom_metrics}, enriched with random polygons
to avoid the risk to bias the study with artificially constructed elements.
Finally, for the solver we considered both the accuracy of the solution and the conditioning of the associated linear system.  The whole framework can be applied to any PEM method, and is highly modular, thus favoring further extensions with new polygons or metrics.

To the best of our knowledge, this work is the first systematic study on the correlation between polygon quality metrics and numerical methods to solve PDEs on discrete polygonal grids. As such, we believe 
it will have great practical impact on two communities. 
For the geometry processing community, metrics with a good correlation with the solver could be incorporated into new meshing tools, producing polygonal grids and refinement/coarsening operators that leverage all the flexibility granted by PEM, thus opening the door to new PDE-aware mesh processing tools. For the math community, correlations will suggest directions to further improve the theory, developing more permissive shape-regularity criteria for the current methods, and new methods to overcome the limitations of the current solvers. 

The outline of the paper is as follows. In Section~\ref{sec:related}, we recall classical results concerning shape regularity criteria for FEM, and discuss how these concepts are addressed in the PEM framework. 
In Section~\ref{sec:benchmark} we provide details on the structure and organization of the benchmark. In Section~\ref{sec:results} we use the benchmark to study a PEM solver based on the Virtual Element Method, and discuss our findings. In Section~\ref{sec:conclusions}, we suggest some directions of future research. Finally, in Appendix~\ref{sec:vem}, we briefly recall the definition and the main properties of the specific PEM solver we used for our numerical experiments.



\begin{figure}[t]
\centering
\includegraphics[width=.7\columnwidth]{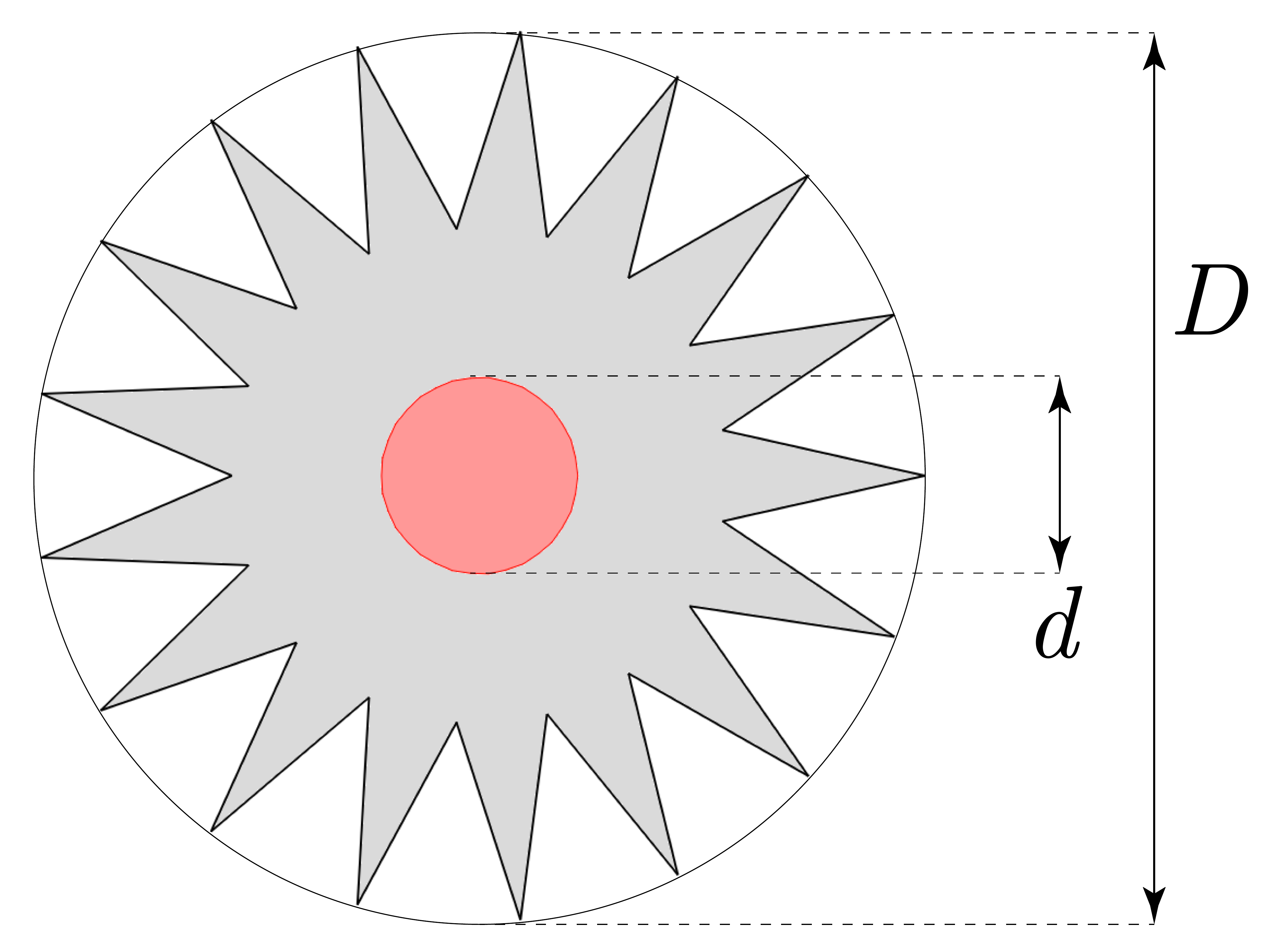}
\caption{Minimal polygon shape regularity is currently expressed in terms of the ratio between the maximal ball inscribed in the kernel of the polygon ($d$), and the maximal ball inscribing the element ($D$).}
\label{fig:shape_reg}
\end{figure}


\section{Related works}
\label{sec:related}




As already mentioned in the introduction, the influence of the shape of the elements on the classical finite element method is quite well understood~\citep{ciarlet2002}. In particular the interested reader can refer to \citep{BRANDTS20082227}, where the equivalence between different shape regularity criteria for triangular and tetrahedral elements is studied.

The PEM family already counts quite a few methods,
such as 
Mimetic Finite Differences~\citep{book_mimetic,brezzi_mimetic}, Discontinuous Galerkin-Finite Element Method (DG-FEM)~\citep{Rev_DG,Cangiani_hpDGVEM}, Hybridizable and Hybrid High-Order Methods~\citep{Cockburn_LDG,Ern_LDG}, Weak Galerkin Method~\citep{Weak_FEM}, BEM-based FEM~\citep{BEM_FEM}, Poly-Spline FEM~\citep{schneider2019poly}, and Polygonal FEM~\citep{pol_FEM}, to name a few. 

\paragraph*{Shape-regularity for VEM.}
As a representative example of the PEM family of approaches, 
one may consider 
the Virtual Element Method (VEM)~\citep{basicVEM}, which can be seen as an extension to FEM for handling general polytopal meshes.
Current a-priori estimates for VEM admit only star-shaped polytopes, and connect the approximation error with the ratio between the radius of the circumscribed circle and the radius of the circle inscribed in the kernel of the element (Figure~\ref{fig:shape_reg}). Moreover, the optimality of the method is only achieved in an even stricter framework (see Assumption~\ref{strong-shape-regularity} in Appendix~\ref{sec:vem}).

\begin{figure*}[t]
\centering
\includegraphics[width=\linewidth]{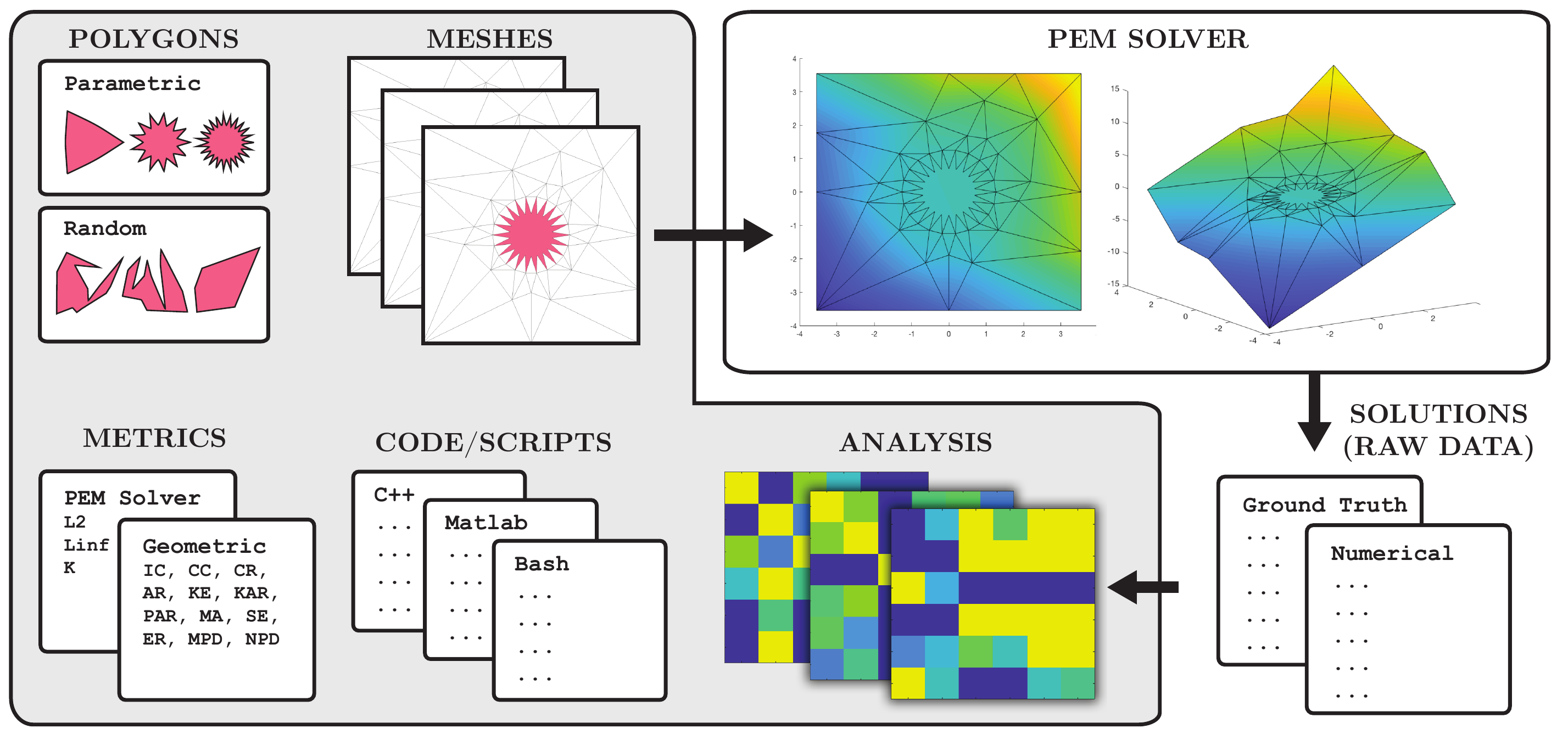}
\caption{Our benchmark (gray shaded area) and how it relates with the PEM solver.}
\label{fig:benchmark}
\end{figure*}

\paragraph*{PDEs in Computer Graphics.}
Solutions of differential equations have been extensively used in computer graphics for a variety of applications, such as mesh parameterization~\citep{floater2005surface}, computation of diffusion distances~\citep{patane2016star}, Voronoi diagrams~\citep{herholz2017diffusion}, and smoothing~\citep{desbrun1999implicit,Liv17_extended}. However, the problem of assessing the capabilities of a numerical method when paired with a particular mesh has been considered only in very recent years.
Restricting to the CG community, the work that is closely related to us is probably~\citep{gao2017evaluating}, where a study on the FEM performances on hexhaderal meshes according to various geometric metrics is proposed. For this work, the authors could leverage both a rich database of hexmeshes produced with various approaches (subsequently released in~\citep{BTPLC18}), and well established quality metrics for hexahedral cells~\citep{stimpson2007verdict}. The benchmark we propose can be seen as its extension to the PEM case. However, this extension is far from trivial, mostly because for the polygonal case the are no available resources in terms of mesh databases, and no consensus has yet been reached on what are the right metrics to consider to evaluate the mesh quality with respect to the performances of the solver.
Schneider et al.~\citep{schneider2018decoupling} identify in mesh resolution, element quality, and basis order the three factors that affect the accuracy of standard FEM, and advocate the use of higher order basis to compensate badly shaped elements. PEM approaches promise to be robuster than FEM against poor elements (Figure~\ref{fig:snowflake}), and the benchmark we propose aims to study how much geometric freedom one can gain, for a fixed basis order. 
Reberol and Levy~\citep{reberol2018computing} propose an efficient voxel-based method to compare two solutions computed on alternative meshes of the same object. In our benchmark ground truth and numerical solution are always computed in the same mesh, therefore sophisticated techniques to transfer a solution from a mesh to another are not necessary.


\section{Benchmark}
\label{sec:benchmark}
We present here the core of our contribution, that is a full benchmark that allows to investigate the correlation of a set of per polygon quality measures ($\SG$) with the performances of a PEM method ($\SP$). The relationship between these entities is computed on a carefully crafted set of discrete polygonal meshes ($\SM$). The main components of the benchmark are therefore the three sets $\SG, \SP, \SM$, the details of which are described in Sections~\ref{sec:polygon_metrics},~\ref{sec:solver_metrics} and~\ref{sec:dataset}, respectively. 

Note that the tool we offer is aimed to be quite general, and can be used to evaluate potentially any PEM solver. To this end, we point out that the solver is not part of the benchmark, but is rather seen as a black box that takes in input a mesh, and returns the solution of a PDE, as depicted in
Figure~\ref{fig:benchmark}.

The major output of the benchmark is a set of measures, both related to the mesh itself and to the PDE solver, computed on the meshes $\SM$. Output data is provided to the user both as raw ASCII files and in statistical format. Statistical data amounts to a correlation analysis between the polygon quality measures in $\SG$ and the performances of the solver, measured according to the metrics in $\SP$. Technical details on how the statistical analysis is performed are given in Section~\ref{sec:correlation}. Such analysis comes in the form of a standard color-coded square matrix (Matlab scripts to load the data and compute the analysis are included in the benchmark). 


As an additional result, the benchmark also outputs correlation analyses between the geometric metrics in $\SG$, and between the performance estimators in $\SP$. Especially for the geometric side, these correlations emphasize any possible redundancy in the set of metrics considered in the study, and therefore offer multiple optimization choices for the design of new meshing tools. To better understand this concept, let us assume that two geometric metrics, $m_1$ and $m_2$, have a strong correlation to each other, and also assume that metric $m_1$ has a strong correlation with the solver performances (e.g., accuracy, stability). Then, meshing algorithms may optimize either for $m_1$ or for $m_2$, obtaining similar (positive) results. Considering that some geometric metrics are much more challenging to optimize than others, having such redundancy is of much practical utility.

For a concrete example of correlation analysis produced with our benchmark, the reader can refer to Section~\ref{sec:results}, where the Virtual Element Method (VEM) is used as  a black box PEM solver. 


\begin{figure}[t]
\centering
\includegraphics[width=.7\columnwidth]{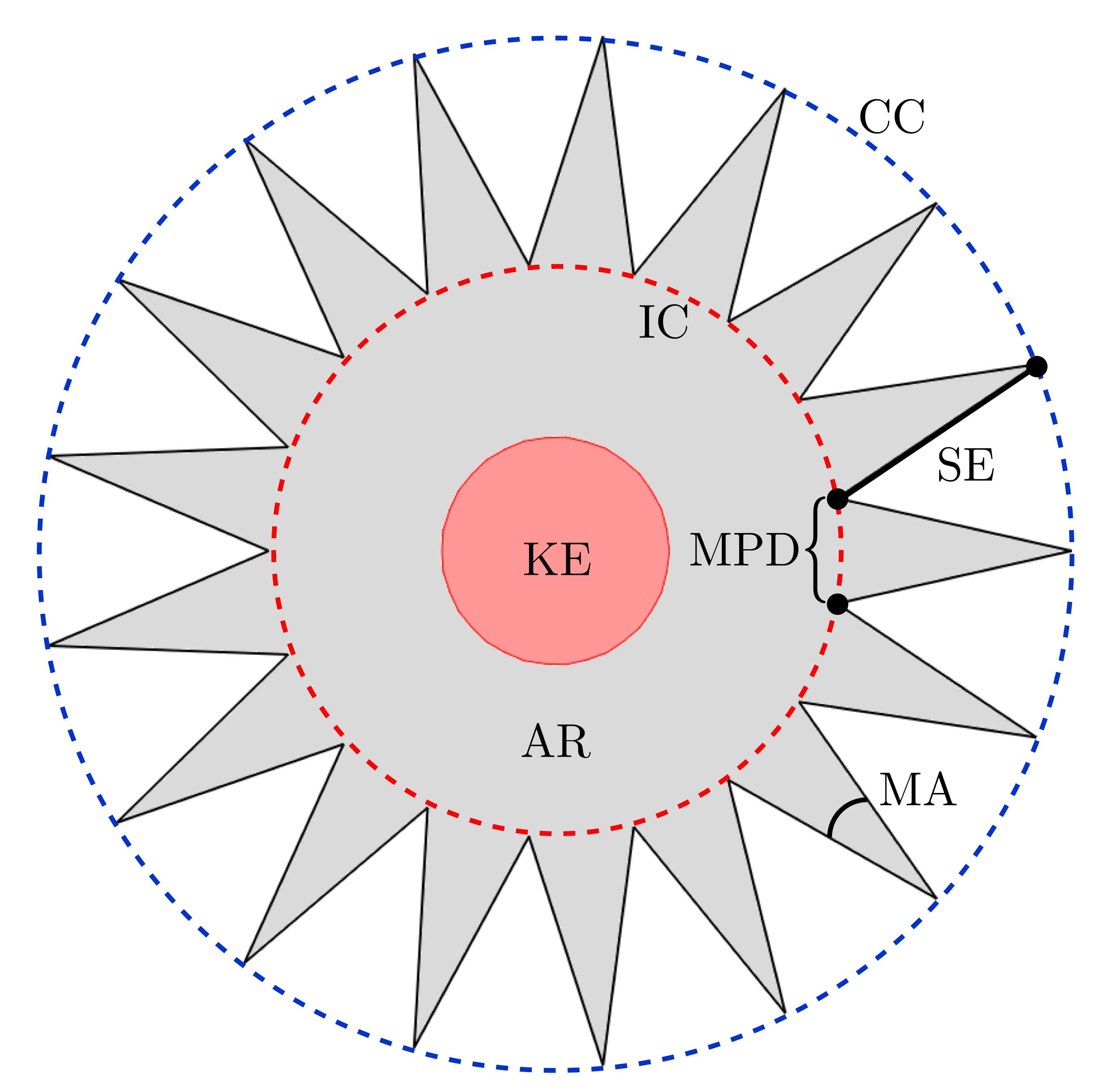}
\caption{Polygon measures we consider in our study: inscribed circle (IC); circumscribed circle (CC); polygon area (AR); kernel area (KE); minimum angle (MA); shortest edge length (SE); and minimum point to point distance (MPD). Relative metrics obtained computing ratios between these quantities are also considered. The full list is available at Table~\ref{tab:geom_metrics}.}
\label{fig:geom_metrics}
\end{figure}

\subsection{Geometric Metrics}
\label{sec:polygon_metrics}
For the geometric part, our study is based on the measures depicted in Figure~\ref{fig:geom_metrics}, which are considered both alone and combined to one another in order to obtain scale invariant metrics. The resulting full list of 12 per polygon metrics is presented in Table~\ref{tab:geom_metrics}. Considering minimum, average, and maximum values of each metric, we obtain a total of 36 per mesh metrics. The benchmark provides C++ code to evaluate these metrics on polygonal meshes, as well as scripts to conveniently run these programs in batch mode on entire datasets of domains. 

Given a polygon $P$, the list of per polygon metrics is the following:

\begin{itemize}
\item \textit{Inscribed Circle (IC):} it measures the radius of the biggest circle fully contained in $P$. Computing the maximum inscribed circle of a general polygon is a though geometric problem. We start from a Voronoi diagram of the edges of $P$, and select as center of the circle the corner in the diagram that is furthest from all edges. The minimum distance between such point and any of the edges of $P$ is the radius of the IC. Note that, differently from the point case, the diagram of a set of segments has curved boundaries between cells. For its computation, we rely on the Boost Polygon Library~\citep{simonson2009geometry};\\

\item \textit{Circumscribed Circle (CC):} it measures the radius of the smallest circle fully containing $P$. We compute it by treating the vertices of $P$ as a point cloud, and the running the Welzl's algorithm to solve the minimum covering circle problem~\citep{welzl1991smallest}. Note that this does not correspond to the classical definition of circumscribed circle, which does not necessarily exist for all polygons (Figure~\ref{fig:cc});\\

\begin{figure}[t]
\centering
\includegraphics[width=.9\columnwidth]{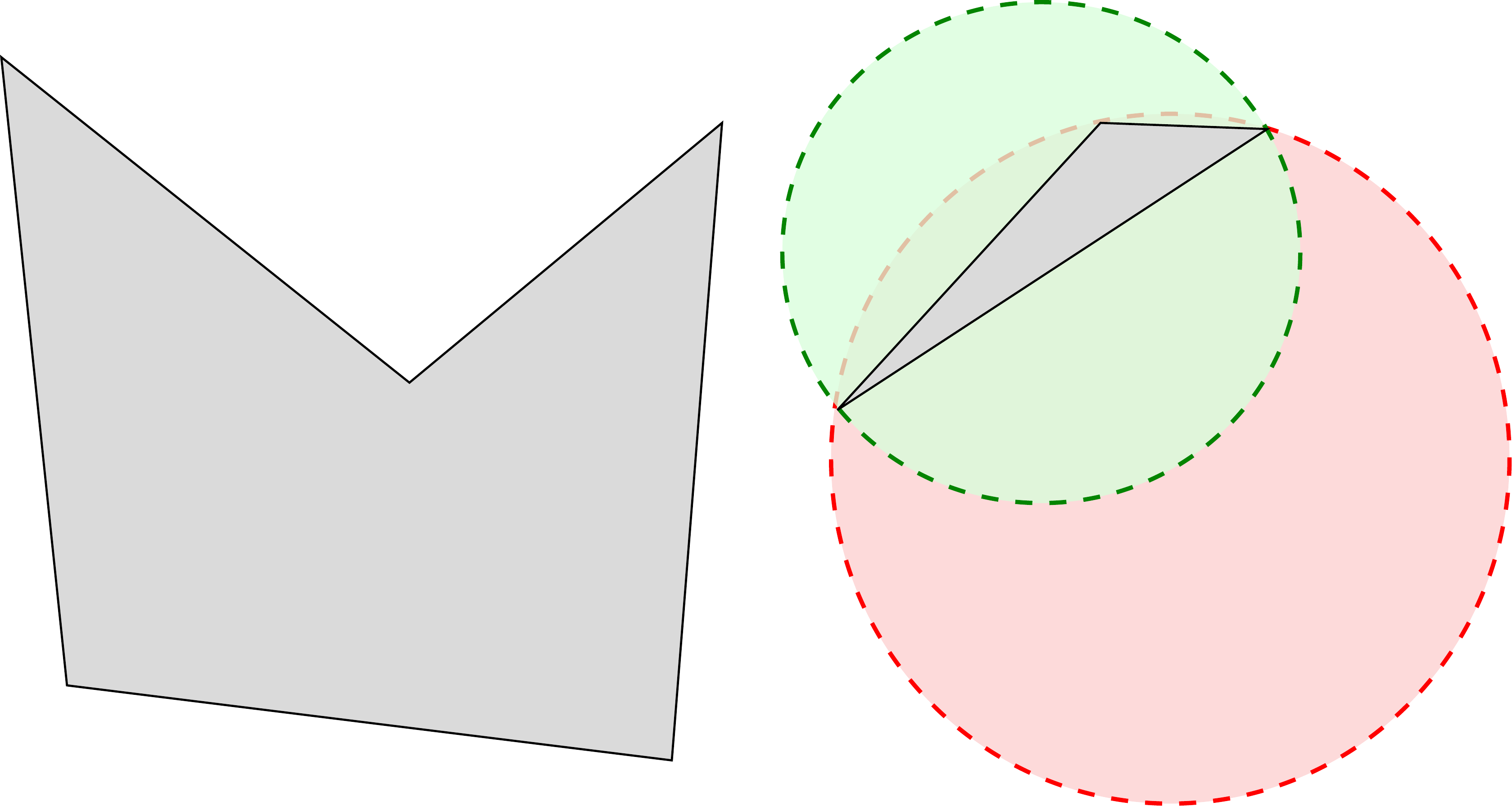}
\caption{Left: a polygon that does not admit a circumscribed circle. To be able to scale on general polygons, we define the radius of the circumscribed circle (CC) as the radius of the smallest circle containing the polygon itself. Right: for a skinny triangle, CC is therefore the radius of the smallest circle that passes through the endpoints of its longest edge (green), and not of the circle passing through all its three vertices (red).}
\label{fig:cc}
\end{figure}

\begin{table}[t]
\centering
\begin{small}
\begin{tabular}{|l|c|l|c|c|}
\hline
\textbf{Metric} & \textbf{Abbr.} & \textbf{Range} & \textbf{Trend} & \textbf{Scale}\\
&&&& \textbf{invariant}\\
\hline
Inscribed circle (radius) & IC & $(0,\infty)$ & -- & No\\
Circumscribed circle (radius) & CC & $(0,\infty)$ & -- & No\\
Circle ratio (IC/CC)& CR & $[0,1]$ & $\uparrow$ & Yes\\
Area & AR &$[0,\infty)$ & -- & No\\
Kernel area & KE &$[0,\infty)$ & -- & No\\
Kernal area ratio (KE/AR) & KAR &$[0,1]$ & $\uparrow$ & Yes\\
Perimeter area ratio & PAR & $(0,\infty)$ & $\downarrow$ & Yes\\
Minimum angle & MA & $(0,\pi)$ & $\uparrow$ & Yes\\
Shortest edge & SE & $(0,\infty)$ & -- & No\\
Edge ratio & ER & $(0,1]$ & $\uparrow$ & Yes\\
Min point to point distance & MPD & $(0,\infty)$ & -- & No\\
Normalized point distance & NPD & $(0,1]$ & $\uparrow$ & Yes\\
\hline
\end{tabular}
\end{small}
\caption{List of 12 per polygon metrics used in our study. The metrics become 36 for a full polygonal mesh (minimum, maximum and averages of each metric are considered). For scale invariant measures, the fourth column indicates whether optimal values are at the top ($\uparrow$) or bottom ($\downarrow$) of the definition range.}
\label{tab:geom_metrics}
\end{table}

\item \textit{Circle Ratio (CR):} it is the ratio between IC and CC. Differently from the previous two, this measure does not depend on the scale of the polygon, and is always defined in the range $(0,1]$;\\

\item \textit{Area (AR):} simply the area of the polygon $P$;\\

\item \textit{Kernel Area (KE):} it is the area of the kernel of the polygon. The kernel of $P$ is defined as the set of points $p \in P$ from which the whole polygon is visible. If the polygon is convex, the area of the polygon and the area of the kernel are equal. If the polygon is star-shaped, the area of the kernel is a positive number. If the kernel is non star-shaped, the polygon has no kernel and KE will be zero;\\

\item \textit{Kernel Area ratio (KAR):} it is the ratio between the area of the kernel of $P$ and its whole area. For convex polygons, this ratio is always 1. For concave star-shaped polygons, KAR is strictly defined in between 0 and 1. For non star-shaped polygons, KAR is always zero;\\

\item \textit{Area Perimeter Ratio (APR):} often referred to as \emph{compactness} of $P$, it is defined as
$$\frac{2\pi*\textsf{area}(P)}{\textsf{perimeter}(P)^2}.$$
This measure reaches its minimum for the most compact 2D shape (the circle), and grows for less compact polygons;\\

\item \textit{Minimum angle (MA):} it is defined as the minimum inner angle of the polygon $P$;\\

\item \textit{Shortest Edge (SE):} it is the length of the shortest edge of $P$;\\

\item \textit{Edge Ratio (ER):} it is the ratio between the shortest and the longest edge of $P$;\\

\item \textit{Minimum point to point distance (MPD):} it is the minimum distance between two (possibly non consecutive) points in $P$. Note that in case the two closest points in $P$ are also consecutive, then MPD and SE are equivalent;\\

\item \textit{Normalized Point distance (NPD):} it is the normalized version of MPD. To make the shortest point to point distance independent from the scale of the polygon, we use the diameter of the minimum circumscribed circle as normalization factor. This bounds NPD within the definition range $(0,1]$.
\end{itemize}

\subsection{Solver performance metrics\label{sec:solver_metrics}}
\label{solver_metrics}
%
To measure the performance of a PDE solver we compute the following quantities, where $u$ and $u_h$ are the ground-truth solution and the solution computed with a PEM solver, respectively:
\begin{itemize}
    \item \emph{Relative $\mathcal{L}_{\infty}$-error}: 
    \[\epsilon_{\infty}:=\|u-u_h\|_{\infty}/\|u\|_{\infty},\]
    where $\|v\|_\infty = \max_{\mathbf x\in\Omega}v(\mathbf x)$;\\

    \item \emph{Relative $\mathcal{L}_{2}$-error}:
    \[\epsilon_{2}:=\|u - u_h\|_{2}/\|u\|_{2},\]
    where $\|v\|_2 = \left(\int_\Omega v(\mathbf x)^2\,d\mathbf x\right)^{1/2}$;\\
    
    \item \emph{Relative error in the discrete energy norm}: \[\epsilon_S:=(\|\mathbf{u}-\mathbf{u_h}\|_S)/\|\mathbf{u}\|_S,\]
    with $\|\mathbf v\|_S = \mathbf v^T\!S\:\mathbf v$, where $S$ is the global stiffness matrix arising from a PEM discretization, and $\mathbf v$ is the vector collecting all the degrees of freedom of $v$;\\
    
    \item \emph{Multiplicative constant, $C$, and estimated convergence rate, $k$}, for the relative $\mathcal{L}_{2}$-error: for most PEM, it can be proved that
    \begin{equation}\label{eq:L2err}
    \epsilon_2 = C h^p,
    \end{equation}
    where $C$ is a constant depending on the shape of the elements of the tessellation, $h$ is the maximum point-to-point distance across all the elements, and $p = 2$. From the previous equation, it follows that
    \[
    \log\epsilon_2 \approx \log C + p \log h;
    \]
    therefore, given a sequence of $M$ mesh refinements, $C$ and $p$ can be estimated using linear regression on the pairs $(\log h, \log\epsilon_2)_i, i = 1, \dots, M$;\\
    
    \item $\mathcal{L}_{1}$ \emph{condition number} \emph{of the stiffness matrix}: $$\kappa_{1}(S) = \|S\|_1\|S^{-1}\|_1,$$ which represents an indicator of the numerical stability of the PEM solver. The higher $\kappa_1$, the less stable the PEM solver. Condition number heavily impacts the performances of iterative solvers, and is therefore important especially when solving big problems that cannot be plugged in a direct solver.
\end{itemize}
%

%
%
%
%

\begin{figure*}
\centering
\includegraphics[height=.95\textheight]{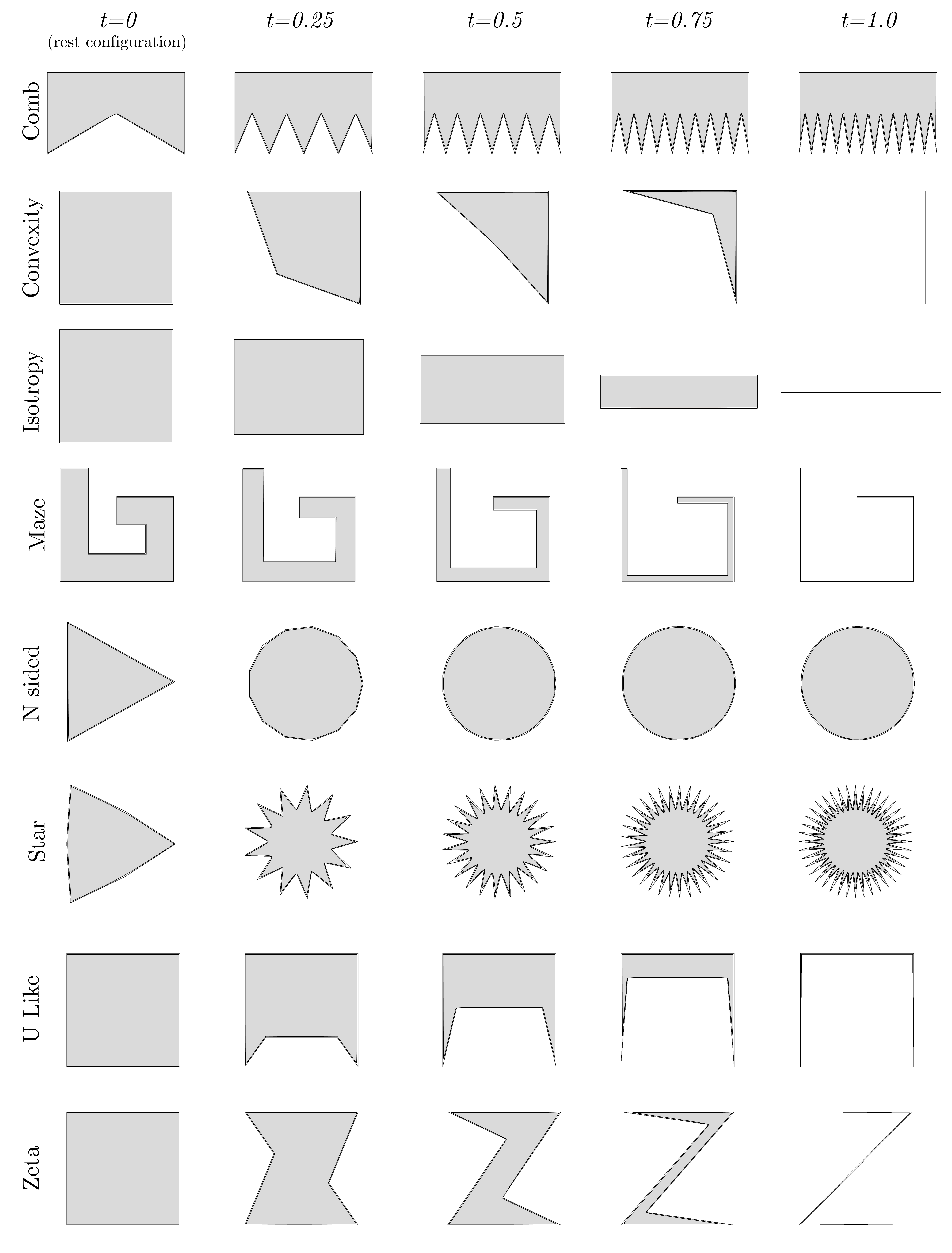}
\caption{The eight families of parametric polygons used to generate the meshes in the dataset. Each polygon starts from a rest configuration ($t=0$, left column), and progressively worsen for growing values of $t$, stressing either one or multiple quality measures listed in Table~\ref{tab:geom_metrics}.}
\label{fig:parametric_polys}
\end{figure*}

\begin{figure*}[t]
    \centering
    \includegraphics[width=\linewidth]{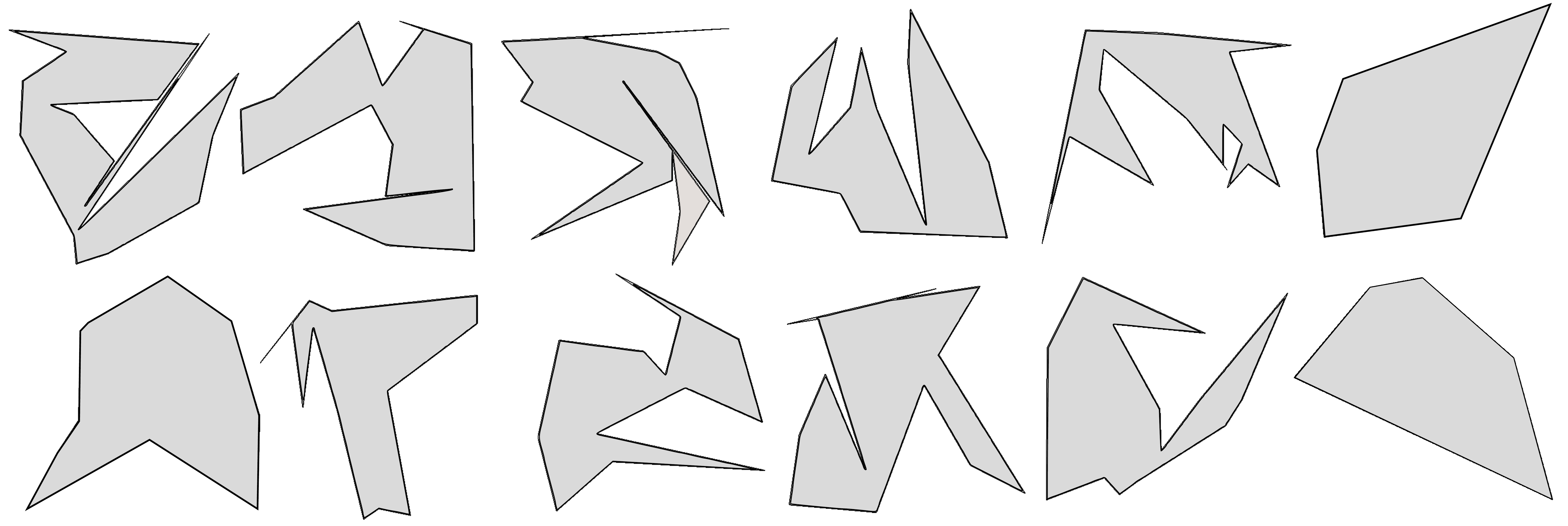}
    \caption{A few examples of randomly generated polygons considered in our benchmark. As can be noticed there is a big variety in terms of complexity of the shapes, spanning from simple convex polygons (top right) to extremely challenging ones (top left).}
    \label{fig:random_polis}
\end{figure*}

\subsection{Polygonal Meshes}
\label{sec:dataset}
In order to evaluate the dependence of the performance of a PEM solver on the geometrical features of the underlying mesh, we propose a set of polygonal meshes explicitly designed to progressively stress the metrics listed in Table~\ref{tab:geom_metrics}. 
We achieved this goal by generating a family of parametric polygons, $P(t)$, with $t \in [0,1]$. Each polygon has a baseline configuration that does not present critical geometric features ($P(0)$), and is progressively made worse by a deformer, controlled by the parameter $t$. We designed 8 different parametric polygons, depicted in Figure~\ref{fig:parametric_polys}. For each polygon, we report both its evolution for growing values of $t$, and the  geometric metrics it affects (Table~\ref{tab:metr_poly_corr}). Note that since the metrics are not independent of one another, each polygon family controls multiple metrics. To minimize any bias due to artificially constructed polygons, we enriched the family of elements with randomly generated polygons, created with CGAL~\citep{cgal:eb-18b}. A few examples are shown in Figure~\ref{fig:random_polis}.

\begin{table}[t]
\centering
\begin{tabular}{|r|c|c|c|c|c|c|c|c|c|c|c|c|}
\hline
& \rotatebox{270}{\textbf{Comb }}
& \rotatebox{270}{\textbf{Convexity }}
& \rotatebox{270}{\textbf{Isotropy }}
& \rotatebox{270}{\textbf{Maze }}
& \rotatebox{270}{\textbf{N Sided }}
& \rotatebox{270}{\textbf{Star }}
& \rotatebox{270}{\textbf{U Like }}
& \rotatebox{270}{\textbf{Zeta }}\\
\hline
\textbf{IC}  & \checkmark & \checkmark&\checkmark & \checkmark& & &\checkmark & \checkmark\\
\hline
\textbf{CC}  & & & & & & & &\\
\hline
\textbf{CR}  & \checkmark & \checkmark & \checkmark&\checkmark & & &\checkmark &\checkmark\\
\hline
\textbf{AR}  & \checkmark & \checkmark & \checkmark& \checkmark& & \checkmark&\checkmark &\checkmark\\
\hline
\textbf{KE}  & & \checkmark & \checkmark& & & &\checkmark &\checkmark\\
\hline
\textbf{KAR}  & &\checkmark & \checkmark& & & \checkmark&\checkmark &\checkmark\\
\hline
\textbf{PAR}  & \checkmark &  & &\checkmark & & & \checkmark&\checkmark\\
\hline
\textbf{MA}  & \checkmark& \checkmark & & \checkmark& &\checkmark &\checkmark &\checkmark\\
\hline
\textbf{SE}  & & &\checkmark &\checkmark & \checkmark& & &\\
\hline
\textbf{ER}  & & &\checkmark &\checkmark & & & &\\
\hline
\textbf{MPD}  & \checkmark & \checkmark &\checkmark & \checkmark &\checkmark &\checkmark & \checkmark&\checkmark\\
\hline
\textbf{NPD} & \checkmark &\checkmark &\checkmark & \checkmark& \checkmark& \checkmark& \checkmark&\checkmark\\
\hline
\end{tabular}
\caption{Correlation between metrics per polygon and the parametric polygons used to generate the meshes in the benchmark.} 
\label{tab:metr_poly_corr}
\end{table}

Given both the parametric and random polygons, the 2D polygonal domains are created by placing an element at the center of a squared canvas, and filling the rest of the domain with triangles, using~\citep{shewchuk1996triangle}.
Since we are interested in the response of the PEM solver with respect to the parametric and random polygons, we aim to fill the rest of the domain with well shaped elements. We obtain the desired effect by imposing $20^\circ$ as minimum inner angle for each triangle. Note that quality constraints are satisfied only away from the central polygon, whereas triangles directly incident to it can be arbitrarily badly shaped, depending on its shape. 

The benchmark contains C++ code to generate single polygonal elements as well as random elements. The program inputs the name of the polygon family, the value of parameter $t$, and a binary flag to enable/disable the generation of the triangular canvas surrounding the polygon. Scripts to conveniently sample the parameter space and create a whole dataset with a single instruction are also included in the software package.

\begin{figure}[h]
\centering
\includegraphics[width=\columnwidth]{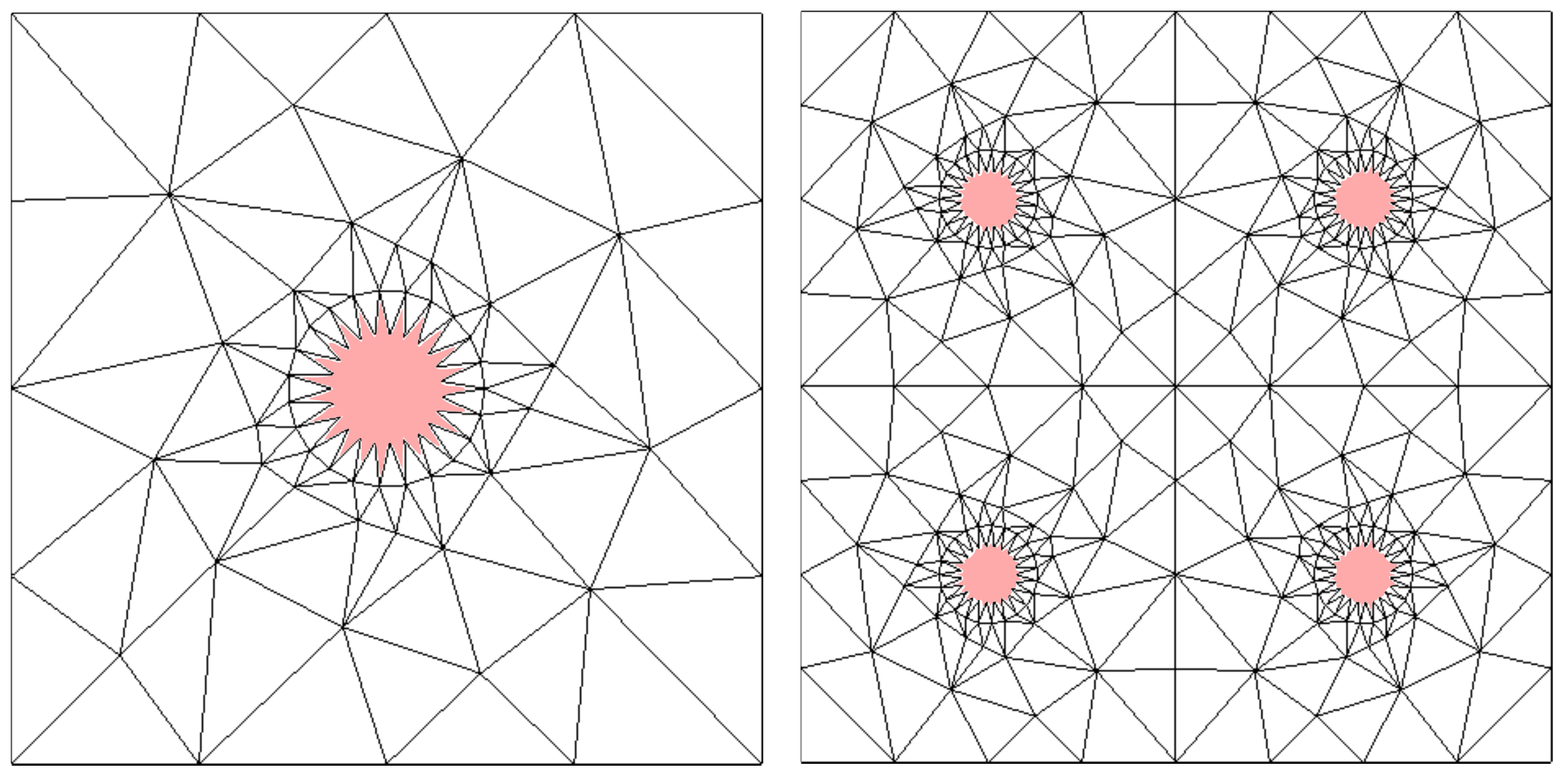}
\caption{An example of 2D polygonal mesh used in the benchmark (left). A mesh hierarchy is also obtained by mirroring the meshes, thus reducing the average edge length and enabling convergence checks (right).}
\label{fig:poly2mirror}
\end{figure}

To complete the mesh generation tools, the benchmark offers the possibility to create a full multi-resolution hierarchy, obtained by mirroring the squared domains multiple times. The users can choose how many levels in the hierarchy they want to create (e.g. to control average edge length and test the convergence of the solver). A few example of polygonal meshes are shown in Figure~\ref{fig:poly2mirror}.


\subsection{Correlation Analysis}
\label{sec:correlation}
To measure the correlation among the geometric metrics and/or performance metrics, the benchmark offers MATLAB\textsuperscript{\textregistered} scripts to load data, measure and visualize the well-known Pearson correlation coefficient (see, for instance,~\citep{DezaDeza}), defined as:
$$
\rho_{a,b}=\frac{\sum_{i=1}^n (a_i -\bar{a}) (b_i -\bar{b})}{\sqrt{\sum_{i=1}^n (a_i -\bar{a})^2 \sum_{i=1}^n (b_i -\bar{b})^2}}
$$
where $a_i$ and $b_i$ are two observations for the mesh $i$ (measures of either geometric or solver quality in our case), $n$ is the number of meshes considered, and $\bar{a}=\sum_1^n x_i$ is the mean of the observations done (analogous for $\bar{b}$).
The correlation coefficient $\rho_{ab}\in [-1,1]$ measures the linear correlation between $a$ and $b$. A value of $1$ implies perfect linear correlation, that is, that there exists a linear equation that perfectly describes the relationship between $a$ and $b$, with all data points lying on a line for which the second observation increases as the first one increases. A value of minus $1$ implies that all data points lie on a line for which the second observation decreases as the first one increases. A value of $0$ implies that there is no linear correlation between these observations.

\subsection{Extensions}
\label{sec:extensions}
The whole benchmark is extremely modular, and its components can be easily extended according to the user needs.
Specifically, all the parametric polygons that populate the set $\SM$ implement the same virtual C++ class. Therefore, extending the family with a new polygon amounts to simply define its base shape, and then implement the function that deforms it according to the value of parameter $t \in [0,1]$. Similarly, new geometric metrics can be added to the system by minimally extending the code-base. For the correlation analysis, MATLAB\textsuperscript{\textregistered} routines to load and process the data are included in the software package, and more complex analysis can be performed leveraging the capabilities of statistical toolboxes.


\section{Benchmarking the Virtual Element Method}
\label{sec:results}
In this section we present the results obtained by coupling the benchmark described in the previous section with a specific PEM solver. Specifically, we used a MATLAB\textsuperscript{\textregistered} implementation of the lowest order Virtual Element Method (VEM)~\citep{mascotto2018}. For a brief introduction on VEM, the reader may refer to Appendix~\ref{sec:vem} and references therein.

As a preliminary step, we investigate the existence of correlations between different geometric metrics (Section~\ref{sec:geogeo}). Subsequently, we consider the correlation between geometric and solver performance metrics (Section~\ref{sec:geosolver}).
For both these studies, we have restricted our investigation to scale-invariant geometric metrics, namely, to CR, KAR, PAR, MA, ER, NPD (see Section~\ref{sec:polygon_metrics} for details). We do not consider scale-dependent metrics because we are interested in the effect of the shape of the elements only (the influence of element size on the performance of the PEM solver is well understood).

\subsection{Test case}

In our numerical tests, we let the computational domain $\Omega$ be the unit square $[0,1]\times[0,1]$, and consider the following model problem
\[
-\Delta u = f,
\]
with Dirichlet boundary conditions and load term $f$ taken so that
$$
u(x,y):=\frac{\sin(\pi x)\sin(\pi y)}{2\pi^{2}}
$$
is the exact solution.
The MATLAB\textsuperscript{\textregistered} VEM black-box prototype returns in output an estimate of the condition number of the stiffness matrix, and of the relative error between the exact and the numerical solution under various norms, as detailed in Section~\ref{sec:solver_metrics}.
This raw data is then processed with the scripts provided by the benchmark, feeding the correlation analysis that aims to study both the accuracy and numerical stability of the solver with respect to the various polygonal meshes generated by the benchmark (Sections~\ref{sec:geogeo} and~\ref{sec:geosolver}), identifying what geometric parameters influence the solver performance.

\subsection{Dataset}
For each of the eight base polygons, we picked 20 equally spaced samples in its parameter domain [0, 1]. This produced 160 different polygons, classified in eight families. Each of these polygons was used to generate a mesh for the unit square domain, as detailed in Section~\ref{sec:dataset}. The 160 resulting meshes constitute our basic dataset $D_0$.
Furthermore, the domain \emph{mirroring} approach described in Section~\ref{sec:dataset} was used to generate four refined versions of $D_0$, that we call $D_1$, $D_2$, $D_3$ and $D_4$, respectively. Note that each $D_i$ has four times more polygons than $D_{i-1}$, and same ratio between number of triangles and general polygons. Following a similar construction, we have also generated a \emph{random} dataset $R_0$ composed of 100 meshes, as well as its associated level of detail mirrored hierarchy $R_1$, $R_2$, $R_3$, $R_4$.
All the aforementioned meshes were generated by using the scripts offered by the benchmark. Bigger (or smaller) datasets can be easily obtained by asking such scripts to sample the space of parametric polygons and the space of random polygons with a different frequency.

\begin{figure}[t]
    \centering
    \includegraphics[width=220pt]{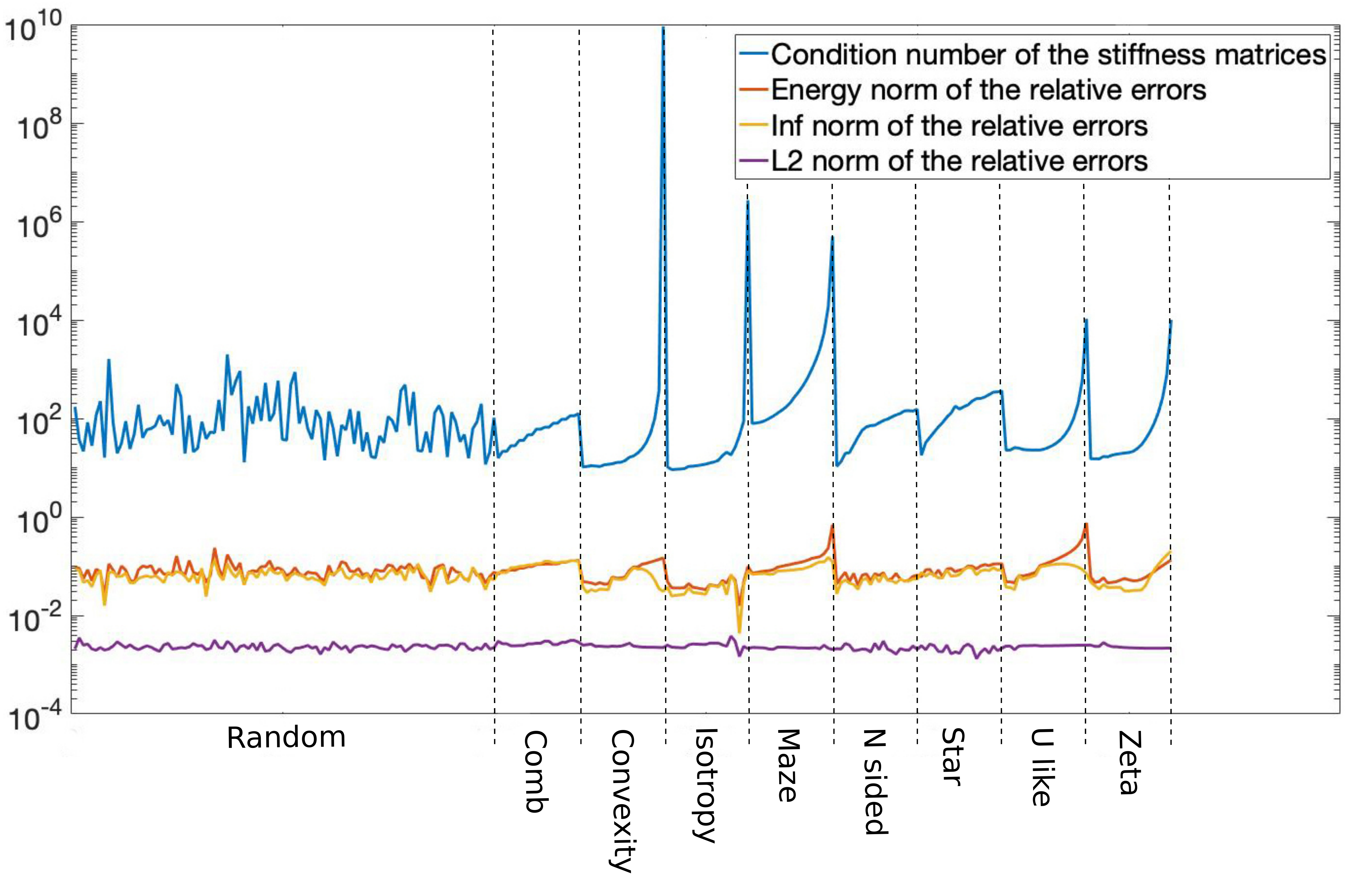}
    \caption{($y$-axis) Solver metrics evaluated on the polygonal meshes ($x$-axis) of the benchmark.}
    \label{fig:HOM-DIRICHLET}
\end{figure}

\subsection{Solver metrics for the benchmark\label{sec:BENCHMARK-METRICS}}

The 6 metrics introduced in Section~\ref{solver_metrics} are used to evaluate different aspects of the PEM solver. This analysis aims at studying the accuracy and numerical stability of the VEM solver defined on the different polygonal meshes of $\Omega$, in order to identify which geometric parameters of the meshes can influence the VEM solver. The results of our analysis are shown in Figure~\ref{fig:HOM-DIRICHLET}.


We observe that the condition number explodes as the quality degenerates for the polygons in the classes \emph{Convexity, Isotropy, Maze, U-like} and \emph{Zeta}, which are the polygons with vanishing area for $t=1$. While this behavior for these classes is to be expected, it is somewhat surprising that the condition number does not explode for the classes \emph{Comb}, \emph{Star} and \emph{N sided}, which lead us to formulate the hypothesis that the metrics ER and NPD have little influence on this performance parameter.

As far as the energy norm of the relative error is concerned, all the classes are affected to different degrees. \emph{Isotropy}, \emph{N sided} and \emph{Star} are affected only minimally, while the most affected are \emph{Maze} and \emph{U like}, where a steep degradation of the error can be observed for very degenerate elements (surprisingly, the error for the \emph{Zeta} class does not seem to present the same behaviour, which deserves further investigation). Particularly interesting are the results for the \emph{Convexity} class, which displays the worst behaviour as far as the condition number is concerned, but where the blowup-like behaviour that we can observe in the two previously mentioned classes, does not occur. This leads us to think that the bad behaviour of these latter classes is indeed resulting from a bad performance of the VEM method itself, and it is not only a consequence of the bad condition number of the stiffness matrix.

\begin{figure*}[t]
    \centering
    \begin{tabular}{cc}
    (a)\includegraphics[height=90pt]{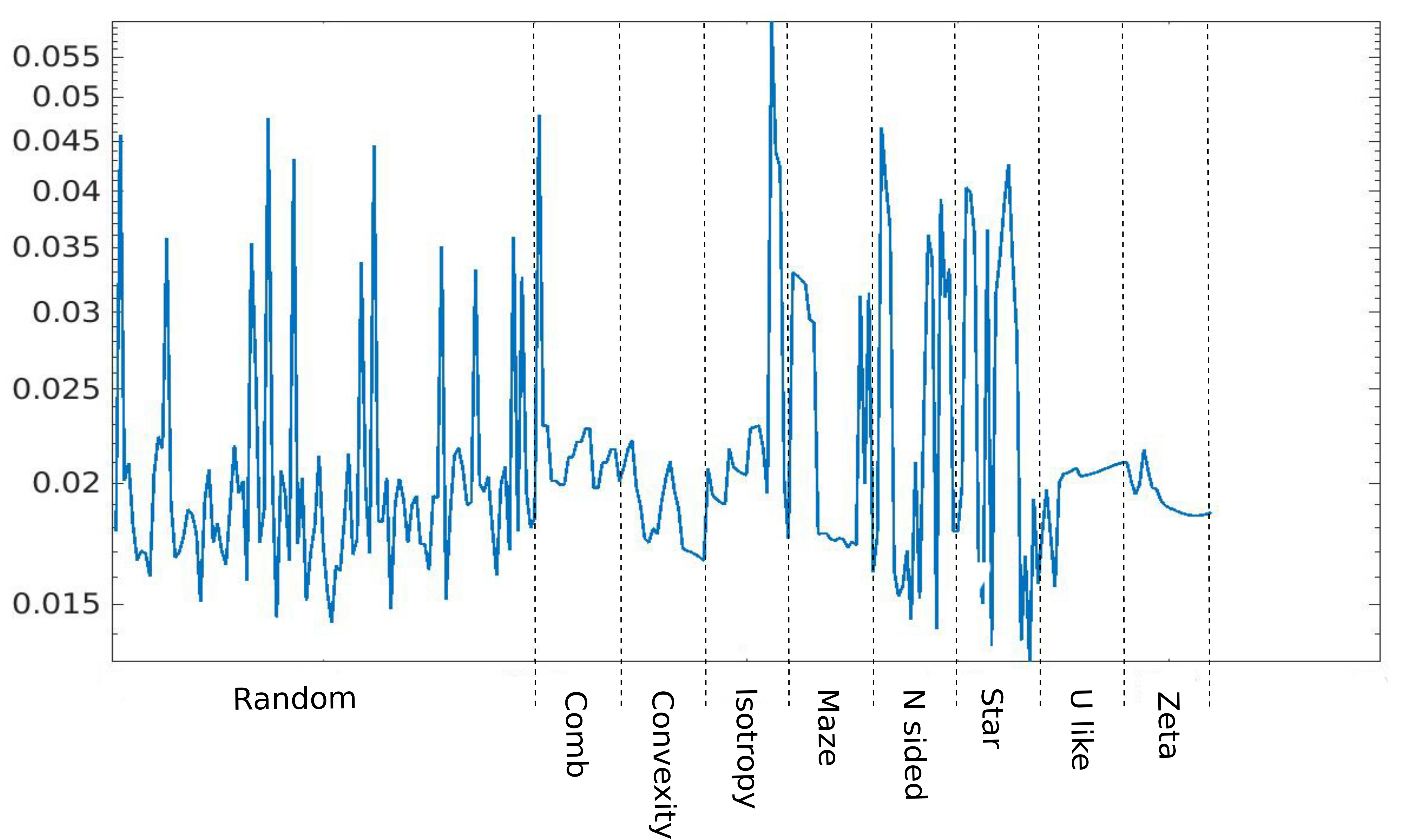}
    &(b)\includegraphics[height=90pt]{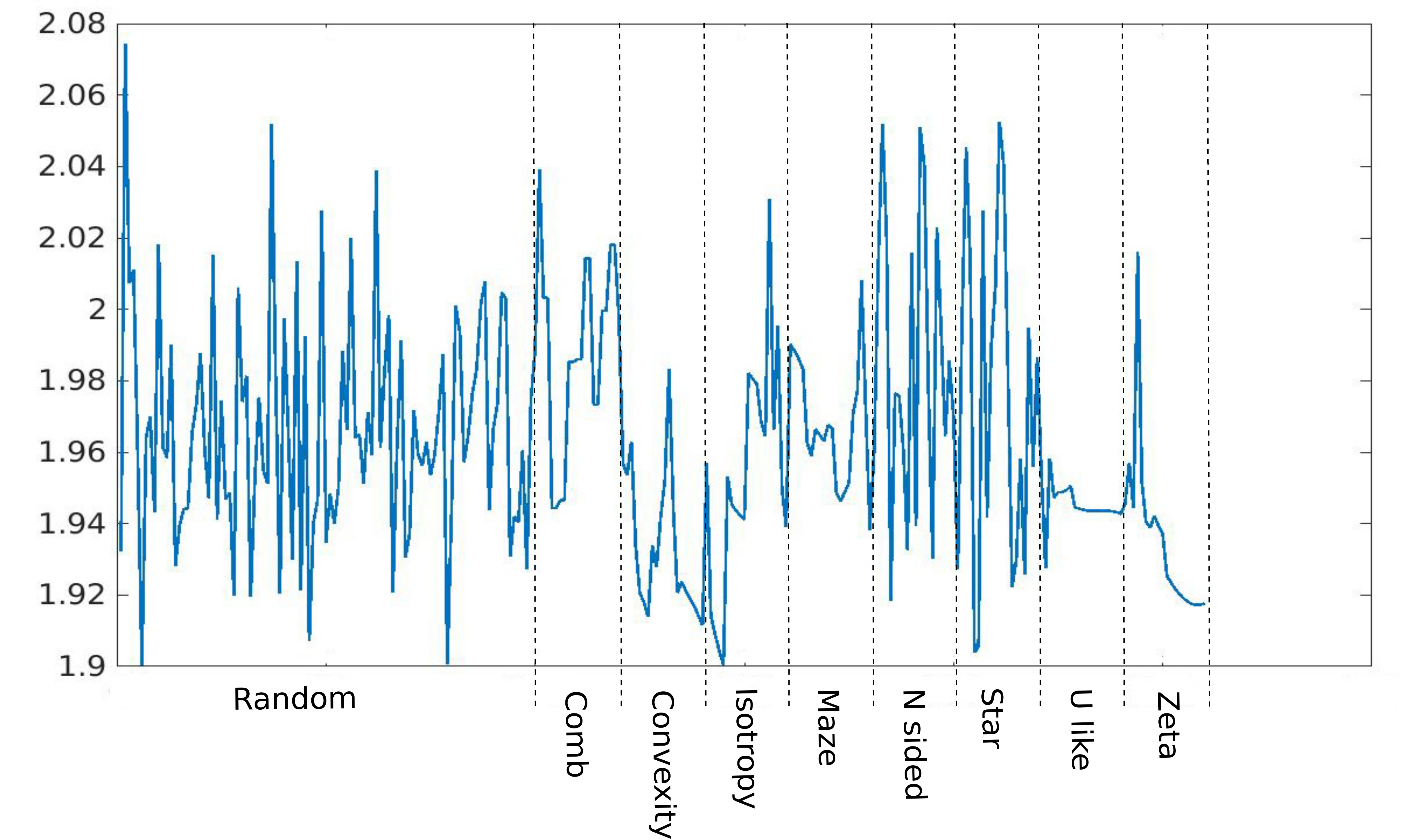}
    \end{tabular}
    \caption{Behaviour of the multiplicative constant of the (a) relative $\mathcal{L}_{2}$-error and (b) estimated convergence rate of $\epsilon_2$.}
    \label{fig:NUMERICAL-ESTIMATES}
\end{figure*}

Finally we used the mirrored data set to estimate the convergence rate $p$ and multiplicative constant $C$ of $\epsilon_2$. The results, shown in Figure \ref{fig:NUMERICAL-ESTIMATES}, are in accordance with the theoretical estimates ($p = 2$) also for the very degenerate cases and confirm the robustness of VEM already observed in the literature.

\subsection{Geometry-geometry correlation}
\label{sec:geogeo}
\begin{figure*}[t]
    \centering
    \includegraphics[width=\linewidth]{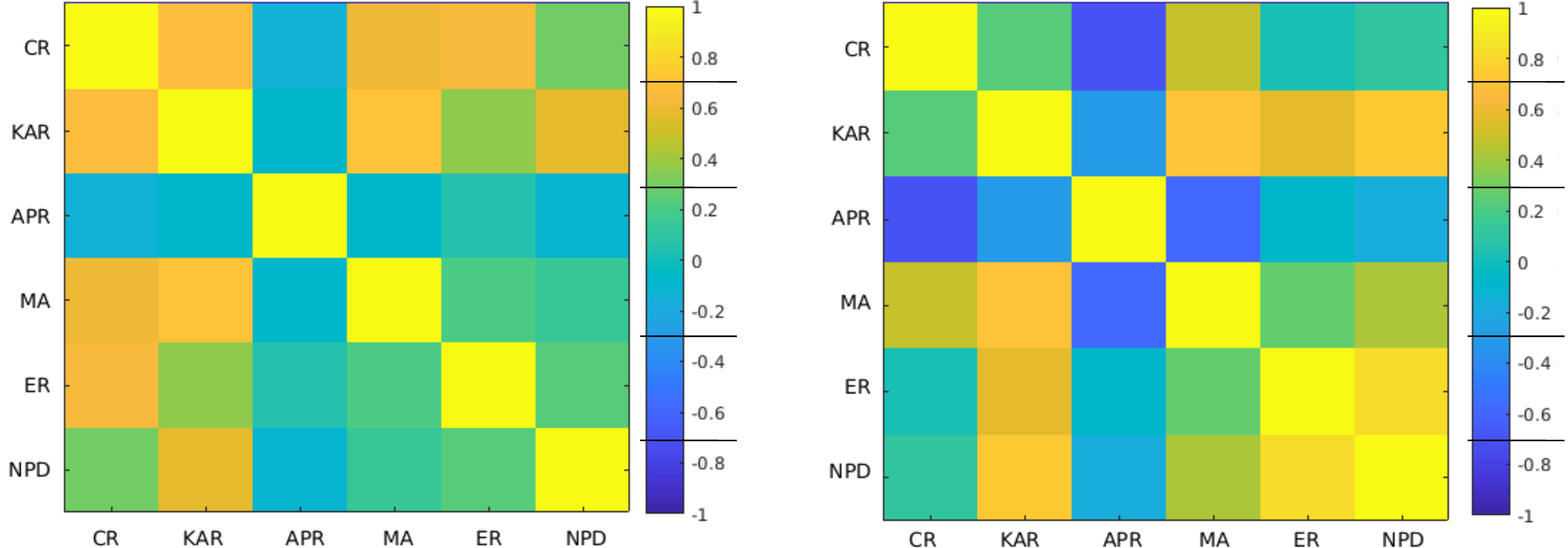}
    \caption{Correlation among the six geometric metrics over the polygon elements of the dataset $D_4$ (left) and over the polygon elements of the random dataset $R_4$ (right). Legend: strong correlation, $\rho \in (0.7,1]$; weak correlation, $\rho \in (0.3,0.7]$; strong inverse correlation, $\rho \in [-1,-0.7)$; weak inverse correlation, $\rho \in [-0.7, 0.3)$.}
    \label{fig:matrix_1_2}
\end{figure*}

Before analyzing the correlation of the different mesh quality metrics with the performance metrics for the VEM method, we start by analyzing the correlations among the different mesh quality metrics. We focus on the non-triangular elements, thus we measure the metrics on these polygonal elements only. Figure~\ref{fig:matrix_1_2} reports the correlation relationship for both $D_4$ and $R_4$, on the left and on the right respectively. As discussed in Section~\ref{sec:correlation}, two measurements positively correlate if $\rho=1$, negatively correlate if $\rho=-1$ and do not correlate if $\rho=0$. When dealing with real data, a threshold is often considered to assess whether two measurements strongly correlate or not. In our setting, we assume $\lbrace 0.7,-0.7\rbrace$ the thresholds for strong positive and negative correlation, respectively, and $\lbrace 0.3,-0.3\rbrace$ the thresholds for weak positive and negative correlation, respectively.
Considering the parametric polygons (Figure~\ref{fig:matrix_1_2}, left) we see that there is not strong correlation among the quality metrics; only a few of them exhibit a moderate positive correlation, while no negative correlation is present.
Stated differently, the selected six metrics are rather independent measures.
Figure~\ref{fig:matrix_1_2} (right) reveals that, when the random polygons are measured, the correlation grows slightly (both in positive and negative sense) with respect to the previous test. This behaviour is not unexpected because random polygons are, on average, farther from the extreme configurations when compared to the parameterized polygons. As a consequence, the measures  cover a smaller range, which induces a higher correlation. Nonetheless, the two matrices in Figure~\ref{fig:matrix_1_2} do not contradict each other.
Finally, we have repeated the same experiments on the refined domains $\{D_{i}\}_{i=1}^{4}$, and their random counterparts $\{R_{i}\}_{i=1}^{4}$. We do not show the results of these experiments as they are essentially identical to those on the unrefined domains. 


\subsection{Geometry-solver correlation}
\label{sec:geosolver}
\begin{figure*}
    \centering
    \includegraphics[width=\linewidth]{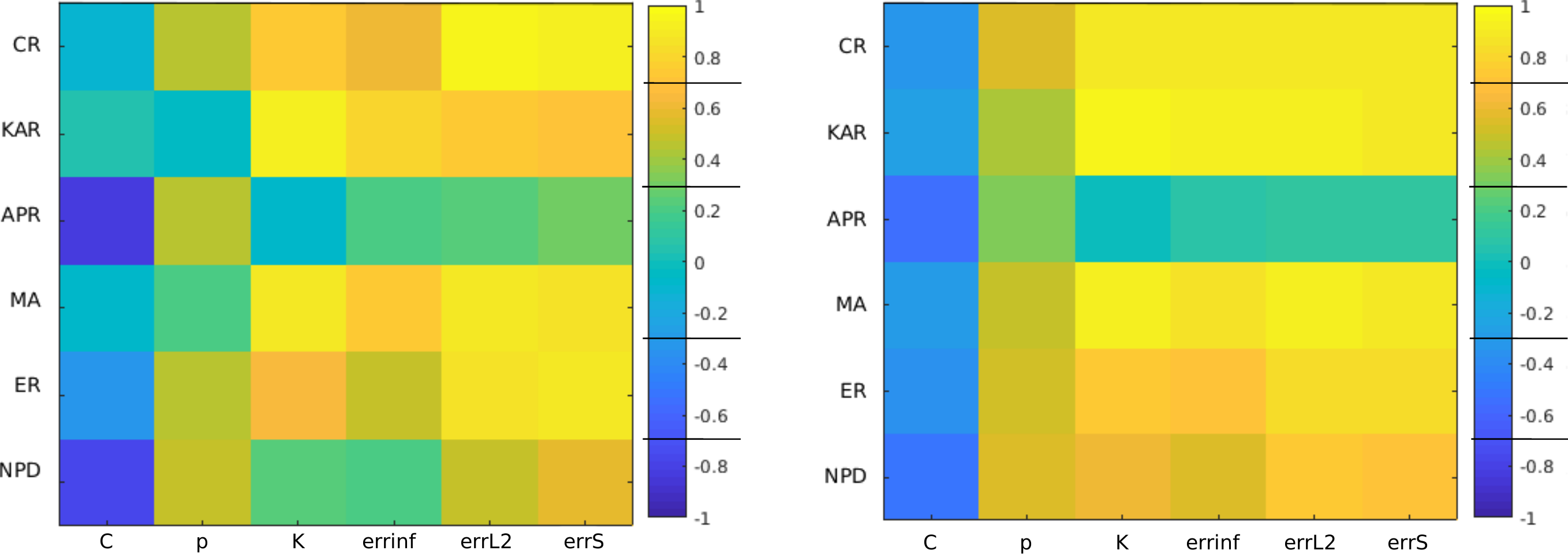}[t]
    \caption{Correlation among the geometric and accuracy metrics for the dataset $D_4$ (left) and $R_4$ (right). Legend: strong correlation, $\rho \in (0.7,1]$; weak correlation, $\rho \in (0.3,0.7]$; strong inverse correlation, $\rho \in [-1,-0.7)$; weak inverse correlation, $\rho \in [-0.7, 0.3)$.}
    \label{fig:matrix_3_4}
\end{figure*}

Even in this case, we have measured the six scale-independent metrics for all the polygonal meshes in $D_4$. However, here we have measured the metrics both on the polygonal element and over the triangular elements in its complement, and we have selected the worst over all these values.
Also, for each mesh we have run our VEM implementation and have compared the solution with the ground-truth using six different error metrics.
By analyzing the correlation between geometric metrics and accuracy metrics on the set $D_4$ we have found that (see Figure~\ref{fig:matrix_3_4}, left side):
\begin{itemize}
    \item KAR and MA are strong positively correlated with $\kappa_{2}(\mathbf{S})$, $\epsilon_{\infty}$, $\epsilon_2$, and $\epsilon_{S}$;
    \item CR is strongly positively correlated with $\kappa_{2}(\mathbf{S})$, $\epsilon_2$, and $\epsilon_{S}$. And, though at a minor extent, with $\epsilon_{\infty}$;
    \item ER is strongly positively correlated with $\epsilon_2$ and $\epsilon_{S}$;
    \item APR and NPD are strongly negatively correlated with $C$.
\end{itemize}

The same experiment was run on our random dataset $R_4$ (see Figure~\ref{fig:matrix_3_4}, right side) and the results, though not contradicting the previous correlations, are less evident due to smaller coverage of the geometric metrics.
As for the geometry-geometry correlation experiment, we have repeated the same tests on the refined domains $\{D_{i}\}_{i=1}^{4}$, and on their random counterparts $\{R_{i}\}_{i=1}^{4}$. The results of these experiments are essentially identical to those on the unrefined domains.

%
%


\section{Conclusions and future works}
\label{sec:conclusions}
We have presented a benchmark to study the correlation between the geometric properties of polygonal meshes, and the performances of Polytopal Element Methods (PEM) in 2D. The benchmark comprises meshes, metrics, code, and various scripts to synthesize, load and process data, and can be coupled with any PEM solver.

To the best of our knowledge this is the first tool to systematically study the correlation between polygonal meshes and PEM solvers. Considering the numerous approaches of this kind recently introduced in literature~\citep{Rev_DG,Cangiani_hpDGVEM,Ern_LDG,Weak_FEM,BEM_FEM,schneider2019poly}, we expect wide adoption of our tool.

In Section~\ref{sec:results} we used the benchmark to evaluate a specific PEM technique: the Virtual Element Method. The results of this study 
confirmed the robustness of the method in terms of the $L^2$ error accuracy, whereas the accuracy in the energy norm is affected by some of the mesh quality metrics. In particular, some specific geometric metrics (e.g. ER, NPD) seem to poorly affect the performances of VEM, whereas a specific class of polygons (i.e. \emph{Convexity}) revealed some weak spot to be further investigated, suggesting the the geometric metrics involved in the evolution of that particular shape, may be a good shortlist of polygon measures to deeply study within the VEM framework, also from a theoretical standpoint.


\subsection{Future works}
As mentioned in Section~\ref{sec:extensions} the benchmark is modular, and can be extended in a number of ways. At PEM level, we are obviously interested in testing the performance of new solvers. For example, we are currently investigating the mutual influence of shape regularity criteria and increasing polynomial degree. Moreover, the results of our analysis, provide us with several interesting research directions relative to the theoretical analysis of the method, such as the study of the effect, or lack thereof, of the presence of small edges on the condition number.

At mesh level, an interesting direction for future improvements consists in studying the space of 2D tilings~\citep{kaplan2000escherization,kaplandihedral}, and try to match it with the concept of parametric polygons used in the benchmark. This would allow to create meshes fully made of polygons subject to study, without having to fill the canvas with triangles. Regarding the correlation analysis, at the moment our study is limited to simple linear correlation. For future versions of the benchmark we plan to extend this part, and introduce more sophisticated (e.g. non-linear) correlation analysis. Finally, we are deeply interested in extending the benchmark to the volumetric case, to study the performances of PEM solvers also on general polyhedral meshes in 3D. Such extension is far from trivial though, as the number of geometric measures and associated metrics to be considered is likely to be considerably higher.
Same goes for the mesh generation phase, where densely sampling the space of parametric 3D polytopes may potentially lead to a combinatorial explosion.



\appendix
\section{\textbf{The Virtual Element Method (VEM)}}\label{sec:vem}

We briefly recall the definition and the main properties of the lowest order Virtual Element Method \citep{basicVEM}. To fix ideas, we focus on the following elliptic model problem:
\begin{equation}\label{eq:poisson}
-\Delta u = f\hbox{ in }\Omega,\qquad u = 0\hbox{ on }\partial\Omega.
\end{equation}
where $\Omega\subset\mathbb{R}^2$ is a polygonal domain and  $f\in L^2(\Omega)$.

We consider a family $\{ \mathcal{T}_h\}_h$ of tessellations of $\Omega$ into a finite number of simple polygons $K$, and let $\Eh$ be the set of edges $e$ of $\Th$. 

\noindent For each polygon $K \in \Th$, let the local space $V^{K}$ be defined as:
\begin{eqnarray*}
	V^{K} = \left \{ v \in H^1(K): v|_{\partial K} \in C^0(\partial K), v|_{e} \in \mathbb{P}_1(e)
	\, \forall e \in \EK, \right . \\
	 \left. \Delta v = 0 \hbox{ in }  K \right\},
\end{eqnarray*}
and the global virtual element space $V_h$ as 
\[
V_h = \{ v \in V: w|_{K} \in V^{K}\ \forall K \in \Th \}.
\]
As degrees of freedom, uniquely identifying a function $v_h \in V_h$, 
we consider the values of $v_h$ at the vertices of the tessellation.

Let $(\cdot,\cdot)$ be the scalar product in $L^2$, 
$a(u,v)= (\nabla u, \nabla v)$ then, 
using a Galerkin approach, we would look for $u_h\in V_h$ such
that for all $v_h\in V_h$ 
\begin{equation}\label{variational}
a(u_h,v_h)=\int_\Omega f \, v_h \, dx. 
\end{equation}
However, both terms at the right and at the left hand side cannot be computed exactly with the knowledge alone of the values of the degrees of freedom of $u_h$ and $v_h$. 
On the other hand, setting, 
for each $K \in \Th$
\[
a^K(u_h,v_h)=\int_K \nabla u_h\cdot \nabla v_h \, dx
\]
we observe that, by using Green's formula, given any $v_h \in V^{K}$ and any $p \in \mathbb{P}_1(K) \subseteq V^{K}$ 
\[
a^K(p,v_h) =  \int_{\partial K} v_h \frac {\partial p}{\partial n}.
\] 

Since on each edge of $K$ $v_h$ is a known linear and $\partial p/\partial n$ is a known constant, the right hand side can be computed exactly and directly from the value of degrees of freedom of $v_h$ and $p$. 
This allows to define  the ``element by element'' exactly computable projection
operator $\Pinabla : V^{K} \longrightarrow \mathbb{P}_1(K)$  
\[
a^K(\Pinabla  u_h, q ) = a^K(u_h,q) \quad \forall q \in \mathbb{P}_1(K), 
\]
and we clearly have 
\[
	a^{K}(u_h,v_h) = a^K(\Pinabla  u_h,\Pinabla  v_h) + a^K(u_h - \Pinabla  u_h, v_h - \Pinabla  v_h).
\]
The Virtual Element method stems from replacing the second term of the sum on the right hand, that cannot be computed exactly, with any computable symmetric bilinear form $\Svem$ satisfying for all $v_h$ with $\Pinabla  v_h=0$
\[
	c_0  a^K(v_h,v_h) \leq  \Svem(v_h,v_h) \leq c_1 a^K(v_h,v_h),
\]
for two positive constants $c_0$ and $c_1$, 
resulting in defining
\[
a^{K}_h(u,v) = a^K(\Pinabla  u,\Pinabla  v) + \Svem(u - \Pinabla  u, v - \Pinabla  v) .
\]

The virtual element discretization of (\ref{variational}) yields the following discrete problem:
\begin{problem}\label{discrete_full} Find $u_h \in V_h$ such that
	\[  a_h(u_h,v_h) = f_h(v_h) \qquad  \forall v_h \in V_h\]
\end{problem}
with 
$a_h(u_h,v_h) = \sum_K a_h^K(u_h,v_h)$ .

\

Problem \ref{discrete_full} is usually analyzed under the following assumption on the polygons of the tessellation.
\begin{assumption}\label{strong-shape-regularity}
	There exist constants $\gamma_0,\gamma_1 > 0$ such that
	\begin{enumerate}[(i)]
		\item  each element $K \in \Th$ is star-shaped with respect to a ball of radius $\geq \gamma_0 h_K$, where $h_K$ is the diameter of $K$;
		\item  for each element $K$  in $\Th$ the distance between any two vertices of $K$ is $\geq \gamma_1 h_K$.
	\end{enumerate}
\end{assumption}
Under this assumption, stability and optimal order one convergence are proven for different choices for the bilinear form $\Svem$ \citep{beirao_stab}), including the simplest one (which we used in the numerical tests)
 of defining $\Svem$ in terms of the vectors of local degrees of freedom as the properly scaled euclidean scalar product.

For details on the implementation 
as well as for the study of the convergence, stability and robustness properties of the method  we refer to \citep{beirao2014hitchhiker,basicVEM}.

Care is needed when computing the relative errors $\epsilon_\infty$ and $\epsilon_2$ for VEM. Since the analytic expression of a VEM function is not known, we can only use its degrees of freedom and all the information we can deduce from them. In particular, for the lowest order VEM, the degrees of freedom are the pointwise values at the mesh vertices $\mathcal P = \{\mathbf p_i\}_i$, $i = 1, \dots, N_\textup{vertices}$. Thus, to compute $\epsilon_\infty$ and $\epsilon_2$, we define
\[
\|u - u_h\|_\infty := \max_i |u(\mathbf p_i) - u_h(\mathbf p_i)|,
\]
and replace $\|u - u_h\|_2$ with $\|u - \Pi^\nabla u_h\|$.

\section*{Acknowledgements}
This work has been supported by the EU ERC Advanced Grant CHANGE, grant agreement No. 694515.



\end{document}